# Industrial Artificial Intelligence

Jay Lee[a], Jaskaran Singh[b], Moslem Azamfar[c,*]

[a] Professor, NSF IUCRC Intelligent Maintenance Systems (IMS), Department of Mechanical Engineering, University of Cincinnati, Cincinnati, OH 45221-0072, USA
Email: lj2@ucmail.uc.edu

[b] Post-Doctoral Fellow, NSF IUCRC Intelligent Maintenance Systems (IMS), Department of Mechanical Engineering, University of Cincinnati, Cincinnati, OH 45221-0072, USA
Email: singh2jn@ucmail.uc.edu

[c,*] Research Scholar, NSF IUCRC Intelligent Maintenance Systems (IMS), Department of Mechanical Engineering, University of Cincinnati, Cincinnati, OH 45221-0072, USA
Email: azamfamm@mail.uc.edu

## Abstract

Artificial Intelligence (AI) is a cognitive science to enables human to explore many intelligent ways to model our sensing and reasoning processes. Industrial AI is a systematic discipline to enable engineers to systematically develop and deploy AI algorithms with repeating and consistent successes. In this paper, the key enablers for this transformative technology along with their significant advantages are discussed. In addition, this research explains "Lighthouse Factories" as an emerging status applying to the top manufacturers that have implemented Industrial AI in their manufacturing ecosystem and gained significant financial benefits. It is believed that this research will work as a guideline and roadmap for researchers and industries towards the real world implementation of Industrial AI.

**Keywords:** Industrial AI, Smart manufacturing systems, Lighthouse factories, Cyber Physical Systems, Industry 4.0

## 1. Introduction

The global race for innovation leadership in smart manufacturing is picking up among companies in Europe, the U.S., and Asia. The manufacturing industry is undergoing unprecedented transformation driven by technologies that help manufacturers to digitize their factories. The fourth

.

industrial revolution can drive financial and operational impact while improving productivity and customer satisfaction. The real-world implementation of Manufacturing 4.0 started with enhanced productivity, followed by improved flexibility, quality, and speed. Manufacturing flexibility can be achieved through machine-to-machine and human-machine interactions in order to form a dynamically changing on-demand production system. Quality improvement can be achieved through real-time plant monitoring and just-in-time maintenance. Degradation of manufacturing equipment and tools diminishes product quality and reduces productivity by increasing unplanned downtime. Therefore, intelligent prognostic and health management (PHM) tools are essential for just-in-time maintenance, which guarantees high-quality products, minimizes unplanned downtime, and increases customer satisfaction. Manufacturing speed can be achieved via increased interconnectivity between different manufacturing sectors contributing to the whole product lifecycle. Cross-company vertical and horizontal data integration can bring transparency and cohesiveness between companies, departments, functions, and capabilities which can significantly increase manufacturing efficiency. Motivated with these short and long term goals, Industry 4.0 realization remains a top priority for many leaders in different manufacturing industries. However, very few manufacturing sites have been able to rapidly adopt these technologies at scale.

Recently, the World Economic Forum, in collaboration with McKinsey & Company, scanned more than 1,000 leading manufacturers in all industries and geographical areas and selected 16 "lighthouses" and recognized them as the most advanced production sites [1]. These so-called lighthouse factories would serve as a beacon for guiding other industries towards the large-scale implementation of M4.0 technologies. We believe organizations require a comprehensive technology roadmap and framework toward adoption, scaling and successful implementation of M4.0 technologies to achieve lighthouse status.

## 2. Evolution of disruptive Manufacturing 4.0 technologies

During the third industrial revolution, the advance of manufacturing technologies relates closely to information technologies such as computer numerical control, flexible manufacturing systems, computer-aided design, computer-aided manufacturing, and computer integrated manufacturing. Achieved at the end of the 20th century, this revolution was characterized by the modernization of computers, the spread of the internet and a wide variety of digital devices. The Internet of Things initiated the fourth industrial revolution, sometimes referred to as Industry 4.0. Even though IoT concepts have been discussed in the literature since at least 1991, it was in 1999 (Figure 1) that Kevin Ashton, a U.K.-based entrepreneur, coined the term '*Internet of Things*' in a presentation to Proctor & Gamble.

In its initial stages, IoT technologies mainly focused on identifying manufacturing elements, assigning an ID to them and improving their connectivity by using information and communication tools. The advent of cloud computing acted as a catalyst for the development and deployment of scalable IoT applications. Cloud computing provided a platform for storage, computation, and communication of data generated by IoT devices. Soon companies were offering software, infrastructure, and platforms as services that in return significantly improved manufacturers' efficiency and reduced costs, while also eliminating the need for companies to develop their own individual communications infrastructures.

In 2009, motivated by advances in these manufacturing technologies, a new paradigm termed as cloud manufacturing emerged. This was touted as a manufacturing version of cloud

.

computing where the cloud provided the core technical support for aggregating distributed manufacturing resources and then extracting and virtualizing them as services. It helped in improving manufacturing efficiency and new revenue stream generation wherein everything in the product lifecycle could be offered as a service. The increasing focus on IoT and cloud manufacturing resulted in an explosive expansion of numerous devices scattered and connected to the Internet, causing production and consumption of manufacturing data at unprecedented levels.

However, the infrastructure and network connections of cloud computing's centralized data center model were not designed to handle the Big Data phenomenon. Therefore, instead of simply extracting raw data from multiple sources and sending it directly to the cloud for analysis, in 2014 Cisco introduced the concept of fog computing – a "highly virtualized platform that provides computing, storage and networking services between devices and cloud computing data centers." Fog computing brings cloud infrastructures close to the edge devices and so improves cloud efficiency by performing edge computation and local storage.

Meanwhile, there have been lots of efforts in integrating cyber and physical systems to improve the integrity of data, flexibility, visualization, and supervisory control. In this regard, a breakthrough five-level Cyber-Physical Production Systems architecture was proposed in 2015 by Dr. Jay Lee from the Center for Intelligence Maintenance Systems (IMS) [2]. The proposed 5C-CPPS architecture provides a clear roadmap for step-by-step implementation of Cyber-Physical Production Systems in Industry 4.0 manufacturing systems. In this design, the digital twin concept is incorporated properly in the cyber level for clarifying the future development in this area. Moreover, blockchain is a promising technology that along with other enablers can significantly improve security, transparency and peer-to-peer interaction at a lower cost and with less energy [3].

A recent White House report on artificial intelligence [4] highlights the significance of AI and the necessity of a clear roadmap and strategic investment in this area. AI is being touted as a major component of this digital transformation. Even with the existing disruptive Manufacturing 4.0 technologies, the majority of connected devices in manufacturing are still not able to make decisions without human intervention, including initiation, management, monitoring, and feedback. Infusing intelligence into these physically connected things can exponentially increase the value that can be generated from them. AI aids the goal of a smart factory; one that would operate with minimal, if any, human interaction.

Building on the impactful 5C-CPPS architecture, IMS has coined a new term: industrial artificial intelligence[4]. It is expected that this concept will enhance productivity, flexibility, quality, and speed in different aspects of manufacturing. Industrial AI can realize smart and resilient industrial systems through four enabling technologies: data technology, analytic technology, platform technology, and operations technology. Supportive technologies like additive manufacturing, augmented reality, and advanced robotics can be a catalyst for speeding up the movement toward M4.0, as well as an adaptive business model to guarantee steady and consistent progress toward new technology implementation and its short and long-term impacts.

## 3. AI as a catalyst to smart manufacturing

The role of AI tools and techniques in smart manufacturing is a hot topic. The AI revolution is beyond its infancy and many companies have significant activity underway. Today more devices

.

– both big and small – deployed on the factory floor are equipped with sensors that gather/share large volumes of data and capture a multitude of actions. Manufacturers have started recognizing the strategic importance of big data analytics and therefore data is becoming a key enabler for enhancing manufacturing competitiveness.

These enormous volumes of data analyzed in real time by leveraging the analytic capabilities of AI can improve decision making and provide enhanced insight to business users - whether that's reducing asset downtime, improving manufacturing efficiency, automating production, predicting demand, optimizing inventory levels or enhancing risk management. PHM is one of the principal applications for the technology, followed closely by demand forecasting, quality control, and robotics. In the last few years, hundreds of venture-backed start-ups have popped up everywhere that are trying to offer AI-coated magic bullets promising to instantly augment enterprise-level insight or help companies to understand a particular machine, process or problem. The winner in this competition would be the one that is going to offer a scalable solution required by an enterprise.

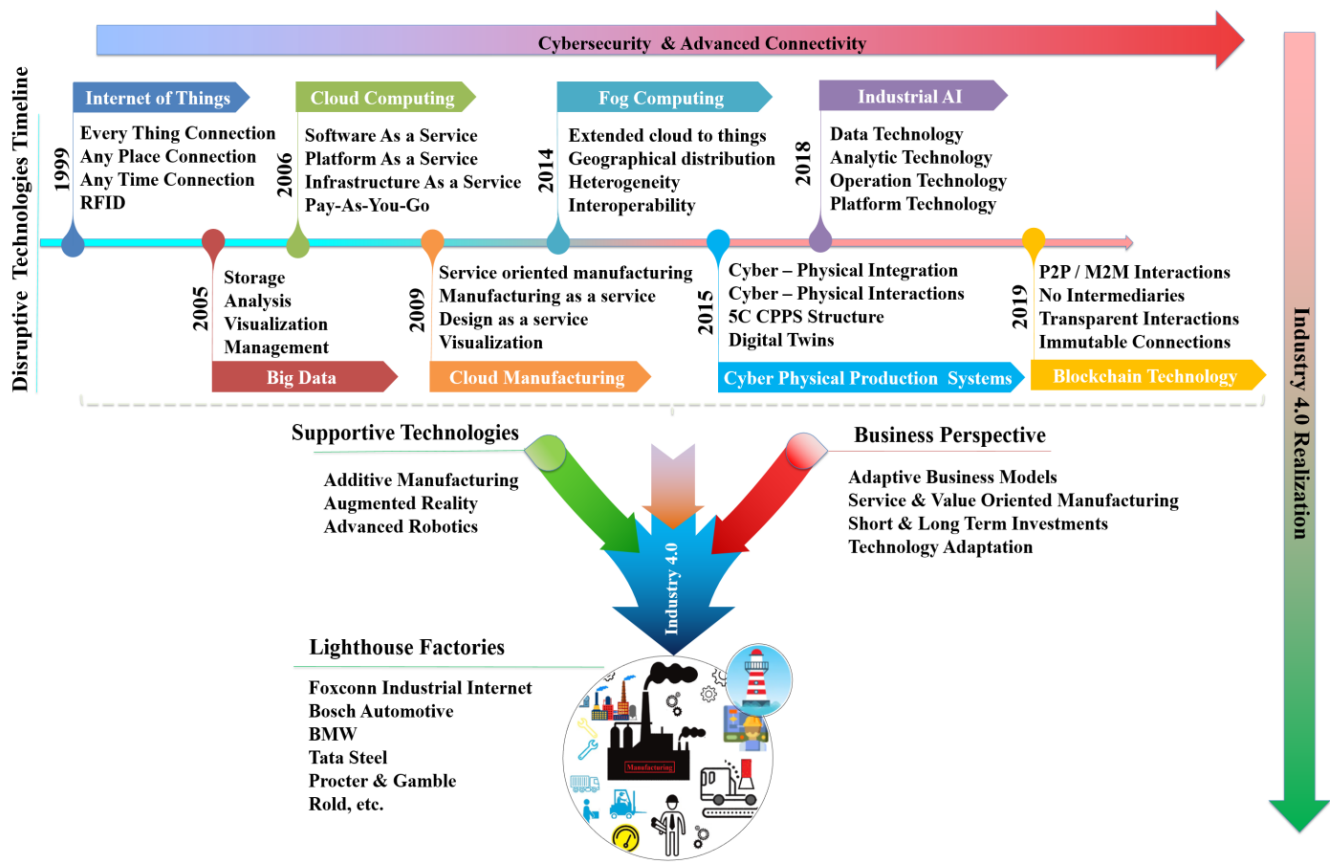

*Fig 1: Evolution of Disruptive Technologies in Manufacturing*

## 4. Digital transformation: stuck in the pilot stage

In order to effectively adapt to this digital transformation, companies need to understand the capabilities of these disruptive technologies and how they can impact the existing ecosystem. However, many companies don't have a comprehensive technology roadmap and framework of
.

how they could embed AI in their strategies and businesses. Hence, they are yet to see meaningful bottom-line benefits from its adoption and thereby most of its economic impact is yet to come. They appear to be stuck in "pilot purgatory" and are not able to succeed in building systematic capabilities.

Automation can be visualized as a batch of robots/machines/equipment performing a set of predefined operations according to a fixed set of rules in a limited number of scenarios. However, AI-based machines don't just follow the rules but intelligently understand and recognize patterns within the data, learn from past experiences and improve future performance. Therefore, AI enables the same set of machines and equipment to identify solutions to complex issues within a given solution space.

While this vision is sound, business leaders all around the world are facing some practical issues hindering real-time deployment of AI specifically in manufacturing industries: 1) Not enough evidence in terms of industrial successes to convince industry to embrace this technology; 2) Lack of a systematic approach for its deployment for a wide range of industrial applications; 3) Non-availability of standardized and structured data from machines/equipment as the data is reported and logged in different formats; 4) Non-availability of machine failure data as industries rarely allow to run their machines to failure; and 5) Dynamic application contexts may require intervention via human operator to add common sense to verify and validate results.

## 5. Industrial Artificial Intelligence can empower smart manufacturing

Companies are in need of a systematic structure for the implementation of AI in industrial environments. Industrial AI can realize smart and resilient industrial systems and enable them to be fault tolerant, on-demand and self-organizing. Industrial AI is defined as "a systematic discipline, which focuses on developing, validating and deploying various machine learning algorithms for industrial applications with sustainable performance." The fundamental concept is the provision of on-demand manufacturing services to end users by optimally coordinating distributed manufacturing resources augmented by AI methodologies. Figure 2 illustrates that industrial AI's four enabling technologies can be better understood when put in the context of the IMS 5C-CPPS architecture. This architecture provides a comprehensive step-by-step strategy from the initial data collection to the final value creation.

.

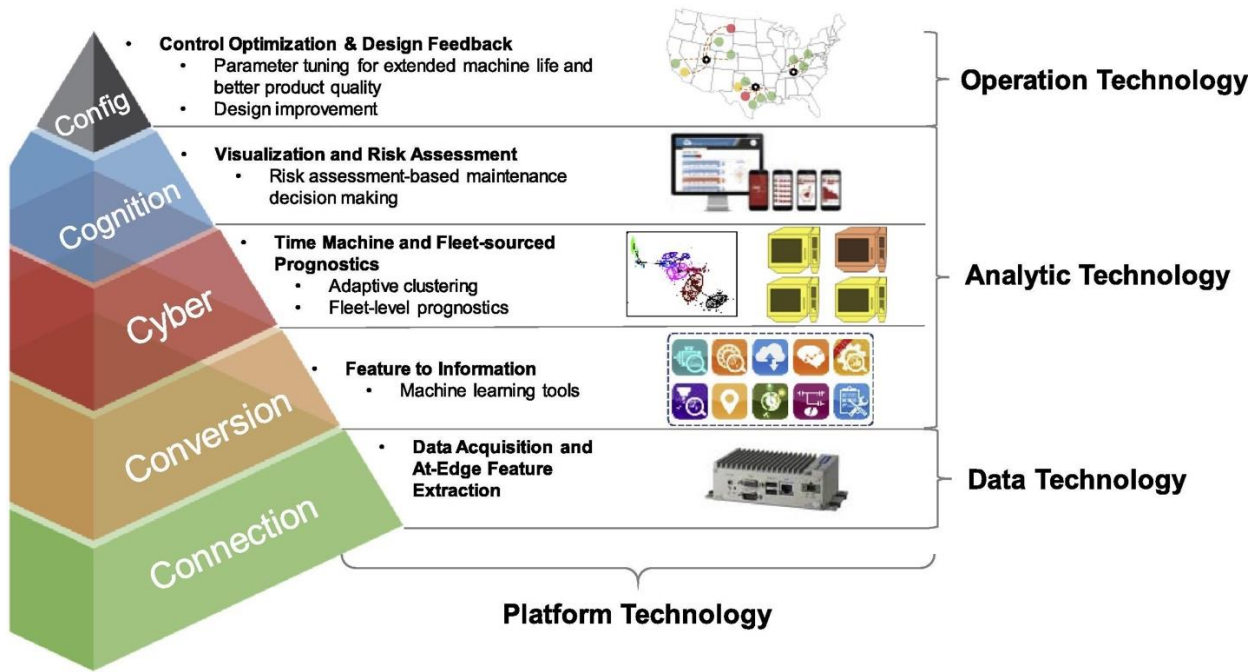

*Fig 2: Enabling Technologies for Realization of CPPS in Manufacturing*

- **Data Technologies**: Data is the proverbial new oil that is fueling the growth of M4.0. Hence, it's critical to understand that the smart factory is primarily about data or more precisely actionable data that leads to information, knowledge, and insights. The more data that is analyzed, the smarter the decisions. Prior to the fourth industrial revolution, there was a heavy reliance on manual methods to collect machine data -- incredibly inefficient, highly susceptible to human error and unable to provide real-time visibility into operations. With a change in the competitive landscape of manufacturing, automating the data collection process from machines and applications is essential to future success.

    Instrumenting the factory floor with sensors can help to create a complete view about the capacity and performance of the company's assets including manufacturing equipment, inventory, people, etc. Different types of signals can be captured from equipment, such as cutting force, vibration, acoustic emission, temperature, current, oil-debris, pressure etc. The data can be static (machine IDs, process parameters, job IDs, etc., i.e. mostly background information) or dynamic (machine health condition, inventory levels, asset utilization, etc.). Data can be generated at either component level, machine level or the shop-floor level and can be broadly divided as structured and unstructured data. The role of data technologies is to efficiently track, control and document this voluminous and varied manufacturing data streaming at high speed in real time. Communication and management of data also become an important part of the data technology for industrial AI. It is expected that the communication technology should have features such as high data transfer rate, low latency, high reliability, high security, accurate traceability, and scalability.

- **Analytic Technologies**: Many companies today have made significant investment in data acquisition hardware and sensors, thus capturing and storing massive amounts of process data. But they typically use them only for tracking purposes, not as a basis for improving operations.

.

It is important to know what to do with the collected information. Analytics refers to the application of statistics and other mathematical tools to these data streams to assess and improve practices. Analytics can enable manufacturers to investigate even the most minute of variabilities in production processes and go beyond lean manufacturing programs such as Six Sigma. It can enable them to segment the production process to the most specific task or activity to identify specific processes/components that are underperforming or causing bottlenecks.

Data-driven modeling can enable manufacturing companies to uncover hidden patterns, unknown correlations and other useful information from manufacturing systems and integrate the obtained information with other technologies for improved productivity and innovation. Apart from data analytics, data visualization tools are an essential element of analytic technologies. The up-to-date real-time information provided by data analytics would be a waste if the results can't be communicated clearly and effectively to the people whose job it is to put them to work. Easy to interpret and user-friendly graphs, charts and reports enable manufacturers to more readily comprehend the analyzed data, track important metrics and assess how far or close they are to their target.

- **Platform Technologies:** Platform refers to both hardware or a piece of software that fills a role in application enablement in an industrial environment, such as connecting devices, handling data (collection/extraction → storage → analyses → visualization) and finally delivering it to the finished applications. Platforms take the center stage in the concept of industrial AI, providing tools and flexibility needed to develop application-centric functions unique to each industry. Platform technologies help in coordinating, integrating, deploying and supporting technologies, such as digital twins, data storage, connectivity across the shop floor, edge intelligence, robotics integration etc. Three major types of platform configurations can be generally found – stand-alone, fog/edge and cloud. The differentiating factor between them is the location where the analytics is deployed.

- **Operations Technologies:** Based on the information derived from the analytics, operations technology, in conjunction with the other technologies, aims to achieve enterprise control and optimization via systems such as product lifecycle management, enterprise resource planning, manufacturing execution systems, customer relationship management, and supply chain management. Finally, outcomes of the analytics performed on the collected data can be fed back to the equipment designer for closed loop lifecycle redesign. In addition, supervisory control and feedback to the physical space are managed through the OT level. The advanced OT incorporated in the manufacturing system is used to form a closed loop management system wherein tasks are generated and executed via intelligent agents running in a distributed and autonomous fashion. OT enables characteristics like self-configure, self-adjust, and self-optimize to the manufacturing ecosystem which finally improve flexibility and resilience throughout the whole production system and lead to higher efficiency and economic impact.

## 6. 'LIGHTHOUSE': A leap forward for realization of smart factories

Manufacturing issues in general can be mapped into two spaces (Figure 3): visible and invisible. Machine breakdown, reduced yield, loss of product quality, etc. are some examples of visible defects. Invisible issues correspond to machine degradation, component wear, lack of

.

lubrication, etc. Generally, well-defined problems such as breakdowns, quality, and productivity etc. are solved through continuous improvement and standard work. i.e. the traditional manufacturing approach (lower left space). Competitive manufacturers today are involving AI algorithms to design, produce and deliver high-quality customer-desired products faster than the competition for problem avoidance (upper left space). Current efforts by many companies have led to the development of new methods/techniques for the invisible issues (lower right space). Adoption of Industrial AI-based approach will aid in producing new value-creation opportunities for smart manufacturing in a dynamic and uncertain environment (upper right space). Unified implementation of the key elements of Industrial AI will not only aid in solving visible problems but avoid invisible ones as well.

Industrial AI can help in achieving the 3W's in smart manufacturing: Work Reduction, Waste Reduction, and Worry-Free Manufacturing. 'Worry' is an invisible concern with today's manufacturing systems that could be due to things like product's bad quality, customer dissatisfaction or business decline. For handling these challenges, advanced AI tools must be utilized through a systematic approach. Work and waste reduction also can be achieved through identifying visible aspects of the problems and addressing their future concerns via the utilization of adaptive AI modules.

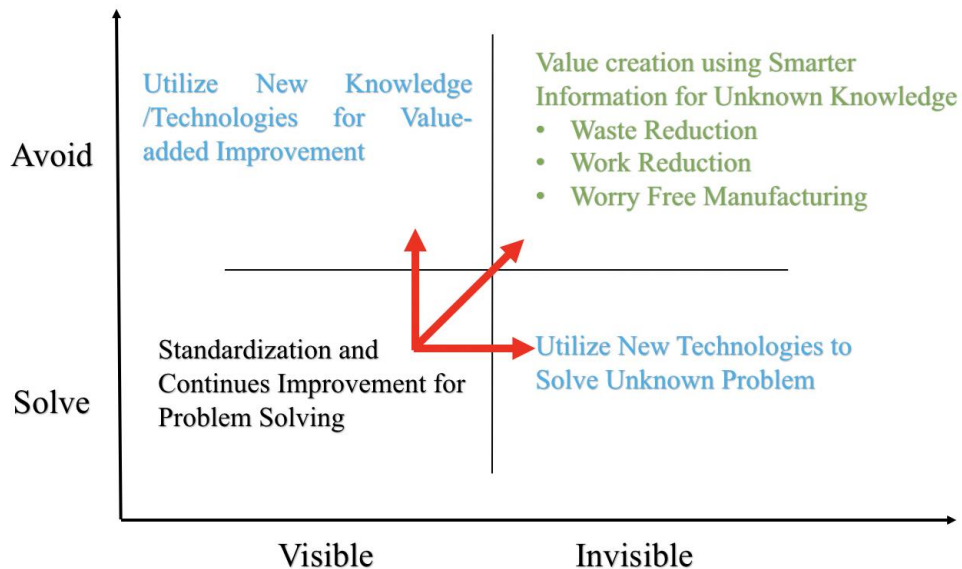

*Fig 3: The Impact of Industrial AI: From Solving Visible Problems to Avoiding Invisible Ones*

Factories that incorporated advanced technologies within their manufacturing development and gained significant economic and financial benefits are the true guiding light in the context of M4.0 and the so-named lighthouse factories. Intelligence is the core asset for such unique factories and is fully utilized through adaptation and implementation of Industrial AI, which can serve as the core driving engine through its four technology enablers and can significantly simplify an advanced CPS structure for its real-world implementation (Figure 4). In this design, manufacturing bottlenecks are identified and managed clearly by autonomous AI modules embedded in the local devices which function collaboratively with each other. For example, a unit of Apple's largest

.

manufacturing contractor Foxconn is one of the lighthouse factories. The company's Shenzhen plant has fully automated most of its production lines and factories, truly becoming a lights-out factory. They are making use of robots in the production line, hence requiring less light – and perhaps more importantly, less human workers. This movement has been followed by other companies, including Bosch Automation, BMW, Tata Steel, Procter & Gamble, and Rold.

*For more information visit: www.imscenter.net*

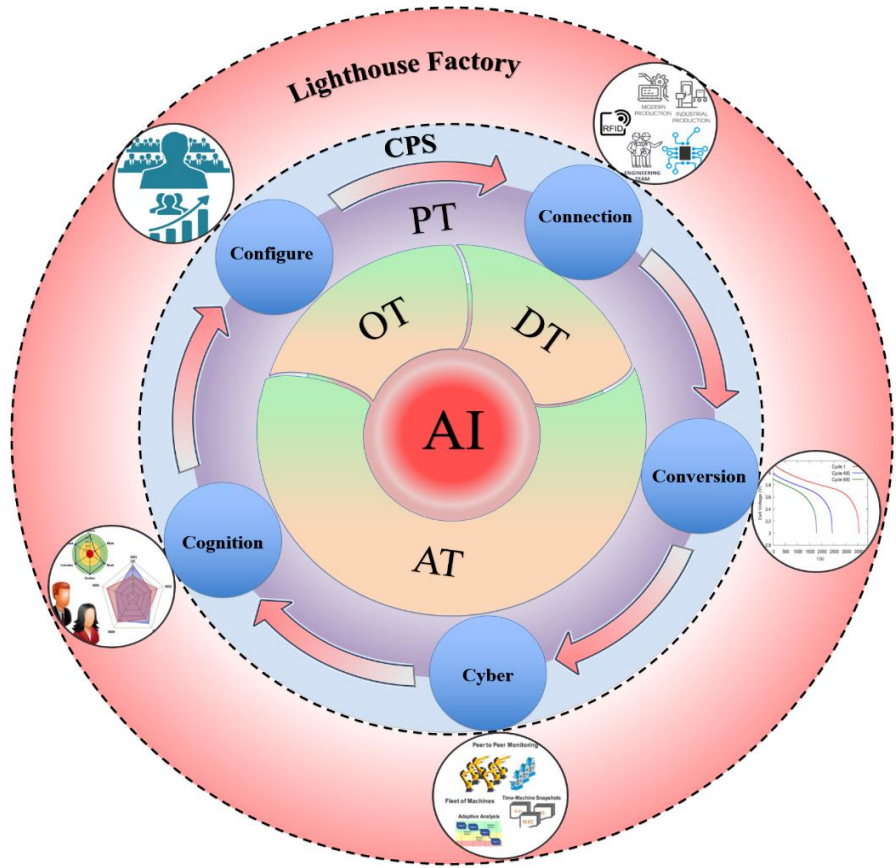

*Fig 4: AI is the Core Asset Driving Growth in Lighthouse Factories*

## 7. Conclusions

This paper presents the key enabling technologies for the realization of industrial AI in manufacturing systems. The key elements of this intelligent system and their functionalities are described in details. Industrial AI is considered as one of the main trends in manufacturing systems and we believe this research would provide guidelines for researches to incorporate this technologies with their developments towards industry 4.0.

.

.